\documentclass{article}

\usepackage{graphicx}

\newcommand{\Tr}{\rm{Tr}}

\begin{document}

\hbox{}
\nopagebreak
\vspace{-3cm}
\begin{flushright}
{\sc OUTP-0013 P} \\
{\sc CPT-P.3998} \\

\end{flushright}

\vspace{1in}

\begin{center}
{\Large \bf Magnetic $Z_N$ symmetry in hot 
QCD and the spatial Wilson loop.}

\vspace{0.5in}
{\large C. Korthals Altes$^1$ and  A. Kovner$^{2}$}\\
\vspace{.4in}
{\small
$^1$Centre Physique Theorique au CNRS, Case 907, Luminy\\
F13288, Marseille Cedex, France\\
$^2$Theoretical Physics, University of Oxford, 1 Keble Road, \\
Oxford, OX1 3NP, UK}\\
\vspace{0.5in}

{\bfseries\sc  Abstract}\\
\end{center}
\vspace{.2in}
We discuss the relation between the deconfining phase transition
in gauge theories and the realization of the magnetic $Z_N$ symmetry.
At low temperature the $Z_N$ symmetry is spontaneously broken while
above the phase transition it is restored. This is intimately related
to the change of behaviour of the spatial 't Hooft loop discussed in
\cite{KAKS}. 
We also point out that the realization of the magnetic symmetry
has bearing on the behaviour of the spatial Wilson
loop. We give a physical argument to the effect that at zero
temperature
the spatial
Wilson loop must have  perimeter law behaviour in the symmetric
phase
but  area law behaviour in the spontaneously broken phase. 
At high temperature the argument does not hold and the restoration of
magnetic $Z_N$ is consistent with area law for the Wilson loop.
\vfill

\newpage
\section{Introduction}

This paper is devoted to further study of theoretical aspects of the
deconfining temperature phase transition in nonabelian gauge theories.
It is an immediate continuation of our earlier work \cite{KAKS}. 
In \cite{KAKS} we showed that the deconfining phase transition in the
pure Yang Mills theory is characterised by the change of behaviour of
the 't Hooft loop operator $V(C)$. In the ``cold'' phase the 't Hooft loop has a
perimeter law behaviour $<V(C)>\propto \exp\{-aP(C)\}$, 
while in the ``hot'' phase it has an area law
behaviour  $<V(C)>\propto \exp\{-\alpha S(C)\}$.

In the present paper we want to sharpen somewhat this observation and further
discuss related questions. We wish to
point out that $V$ is in fact an order parameter which probes the
breaking of a physical symmetry  of the Yang Mills theory.
The symmetry in question is the magnetic $Z_N$ symmetry discussed by
't Hooft \cite{thooft}.
The deconfining phase transition is therefore characterized by
the change in the mode of realization of a global $Z_N$ symmetry:
the symmetry is broken spontaneously in the ``cold'' phase while it is
restored in the ``hot`` phase.

The previous two paragraphs may sound at first like a red
herring. After all an order parameter for the deconfining phase
transition as well as a related $Z_N$ symmetry have been discussed
for many years. The order parameter in question is the free
energy of an external static colour source in the fundamental representation:
the Polyakov line  $P={\rm Tr} P\exp\{ig\int_0^\beta d t A_0\}$.
The $Z_N$ symmetry is the transformation
$P\rightarrow\exp\{i{2\pi\over N}\}P$. We
will refer to this
transformation as the electric $Z_N$.
There is however a great difference between the physical nature of $P$
and $V$ and the associated $Z_N$ symmetries.
The operator $V$ is a canonical operator in the physical Hilbert space of
the Yang Mills theory. 
The magnetic $Z_N$ symmetry
similarly is a transformation that acts on quantum states in the physical
Hilbert space.
On the other hand $P$
has a
very different status. It is not an operator in the Hilbert space and
as such not a canonical order parameter.
It appears as an auxilliary
object when projecting onto gauge invariant physical subspace of the
Hilbert space. 
The ``electric'' $Z_N$ - the operation that
transforms $P$ by multiplying it by a phase - similarly
is not a canonical
symmetry. 
These issues were discussed in detail in \cite{smilga}.
There is no transformation of states in the physical Hilbert
space that is related to this ``symmetry'', although it is indeed a
symmetry of the Euclidean path integral representing the statistical sum.

This is not to say of course that $P$ and electric $Z_N$ are useless concepts.
The standard effective action, defined by the constrained path integral
\begin{equation}
\exp{-S_{eff}(P)}=\int DA_0 \delta(P-P(A_0))\exp{-S(A)}
\label{seff}
\end{equation}
is gauge invariant. 
It is instrumental in computing the vortex expectation value.
The way the electric $Z(N)$ symmetry is realized in $S_{eff}$ is also
related to the behaviour of the order parameter of
the magnetic $Z(N)$. We will discuss this in detail
in section 3 ~\cite{KAKS}. 

However if one wants to describe the deconfinement phase
transition in terms of a canonical order parameter 
in the same way as the Ising transition is described in terms of
magnetisation,  one should zero in on $V$ rather than on $P$ and
should study the magnetic $Z_N$ symmetry rather than electric $Z_N$.
This is what we intend to do in this paper. 

The action of the magnetic $Z_N$ symmetry is very different in 2+1 and 3+1
dimensional cases. In 2+1 dimensions it acts very much like usual global
symmetry in a scalar theory with the order parameter being 
a scalar vortex field. 
In 3+1 dimensions the symmetry acts 
not like a standard global symetry - its ``charge'' is an
integral over a two dimensional spacelike surface rather than over the
whole of the three dimensional space\footnote{These type of symmetries
nowadays are frequently discussed in the context of ``M -
theory''\cite{branes}.}. As a consequence its order parameter
is not a local field but rather a magnetic vortex 
stretching over macroscopic distances.

It is therefore convenient to start the discussion with the three dimensional
gauge theories and to present all the arguments in this case. 
The generalization of appropriate aspects of this discussion 
to 3+1 dimensions will be given 
in the last part of every section.

The plan of this paper is the following. In Section 2 we recap the
definition of the 't Hooft loop and its 2+1 dimensional analog - the
magnetic vortex operator. We formulate the arguments for the existence
of the magnetic $Z_N$ symmetry in theories without fundamental matter fields.
We also show by explicit construction that the generator of this 
symmetry in the pure gluodynamics is none other than the spatial Wilson loop.

In Section 3 we discuss the relation between the behaviour of the
't Hooft loop and the realization of the magnetic $Z_N$ in the ground
state of the theory. We demonstrate that the mode of the realization
of the symmetry changes at the deconfining phase transition, while
spontaneously broken at low temperature the symmetry is restored above
the phase transition.

In Section 4 we present in a toy model a simple physical
picture explaining how the behaviour of spatial Wilson loop
discriminates at zero temperature between the phases with broken
and unbroken magnetic $Z_N$. In the phase  where the $Z_N$ symmetry is
broken, $W$ must have an
area law 
while in the case of unbroken $Z_N$ it must have perimeter law.
We explain why this argument does not generalize to the high
temperature phase and thus why the area law behaviour of the Wilson
loop in the hot phase is consistent with restoration of the magnetic
$Z_N$ symmetry. 

Finally in Section 5 we conclude with a short discussion.

\section{The magnetic $Z_N$ symmetry and the 't Hooft loop operator.}

In this section we discuss the notion of the magnetic $Z_N$ symmetry and
 its order
parameter - 't Hooft loop, or magnetic vortex operator. 
Most of the material contained here
is not new and, perhaps with the exception of explicit
identification of the $Z_N$ generator with the spatial Wilson loop, 
is contained in \cite{thooft,zn,kovner1}.
At the risk of being repetitive we have 
decided nevertheless to include this extended introductory part, 
since we feel that
the concept of magnetic $Z_N$ symmetry is not widely appreciated in
the community. The $Z_N$ symmetry structure 
is the basis
for our discussion of the deconfining phase transition 
in the following sections.

Let us start by recalling the argument due to 't Hooft that a
nonabelian $SU(N)$ gauge theory with charged fields in adjoint representation
posesses a global $Z_N$ symmetry \cite{thooft}. 

We discuss the 2+1 dimensional
case first.
Consider a theory with several adjoint Higgs fields so that varying
parameters in the Higgs sector 
the $SU(N)$ gauge symmetry can be broken completely. In
this phase the perturbative spectrum will contain the usual massive
``gluons'' and Higgs particles. However in addition to that there will
be heavy stable magnetic vortices. Those are the analogs of
Abrikosov-Nielsen-Olesen vortices in the superconductours and they
must be stable by virtue of the following topological argument.
The vortex configuration away from the vortex core has all the fields
in the pure gauge configuration
\begin{equation}
H^{\alpha}(x)=U(x)h^{\alpha}, \ \ \ \ A^\mu=iU\partial^\mu
U^\dagger
\end{equation}
Here the index $\alpha$ labels the scalar fields in the theory,
$h^\alpha$ are the constant vacuum expectation values of these fields, and
$U(x)$ is a unitary matrix. As one goes around the location of
the vortex in space, the matrix $U$ winds nontrivially in the gauge
group. This is possible, since the gauge group in the theory without
fundamental fields is $SU(N)/Z_N$ and it has a nonvanishing first
homotopy group $\Pi_1(SU(N)/Z_N)=Z_N$. Practically it means that when
going around the vortex location full circle, $U$  does not return to
the same $SU(N)$ group element $U_0$, but rather ends up at
$\exp\{i{2\pi\over N}\}U_0$. Adjoint fields do not feel this type of
discontinuity in $U$ and therefore the energy of such a configuration
is finite. Since such a configuration can not be smoothly deformed
into a trivial one, a single vortex is stable. Processes involving
annihilation
of N such
vortices into vacuum are allowed since N-vortex configurations 
are topologically
trivial.
One can of course find explicit vortex solutions once the Higgs
potential is specified. 
As any other semiclassical solution in the weak coupling
limit the energy of such a vortex is inversely proportional to the
gauge coupling constant and therefore very large.
One is therefore in a situation where the spectrum of the theory
contains a stable particle even though its mass is much higher than
masses of many other particles (gauge and Higgs bosons) 
and the phase space for its decay into
these particles is enormous. The only possible reason for the existence of
such a heavy stable particle is that it must carry a conserved quantum
number.
The theory therefore must possess a global symmetry which is unbroken
in the completely higgsed phase. The symmetry group must be $Z_N$
since the number of vortices is only conserved modulo $N$.

Now imagine changing smoothly the parameters in the Higgs sector so
that the expectation values of the Higgs fields become smaller and
smaller, and finally the theory undergoes a phase transition into the
confining phase. One can further change the parameters so that the
adjoint scalars become heavy and eventually decouple completely from
the glue. This limiting process does not change the topology of the gauge
group and therefore does not change the symmetry content of the
theory. We conclude that the pure Yang-Mills theory also
posesses a $Z_N$ symmetry. 
Of course since the confining phase is separated from the completely
Higgsed phase by a phase transition one may expect that the $Z_N$
symmetry in the confining phase is represented differently. In fact
the original paper of 't Hooft as well as subsequent work\cite{zn}
convincingly 
argued that in the confining phase the $Z_N$ symmetry is spontaneously
broken and this breaking is related to the confinement phenomenon.

The physical considerations given above can be put on firmer formal
basis. In particular one can construct explicitly the generator of the $Z_N$
as well as the order parameter associated with it 
- the operator that creates the
magnetic vortex \cite{kovner1}.
We will now describe this construction. 

\subsection{The Abelian case}

Consider first an Abelian gauge theory.
In this case the homotopy group is $Z$ and therefore we expect the
$U(1)$ rather than $Z_N$ magnetic symmetry. It is in fact absolutely
straightforward to identify the relevant charge. It is none other than
the magnetic flux through the equal time plane, with the associated
conserved current being the dual of the electromagnetic field strength
\begin{equation}
\Phi=\int d^2x B(x),\ \ \ \ \ \ \ \ \ \ \partial^\mu\tilde F_\mu=0
\label{phiqed}
\end{equation}
The current conservation is insured by the Bianchi identity.
A group element of the $U(1)$ magnetic symmetry group is
$\exp\{i\alpha\Phi\}$ for any value of $\alpha$.
A local order parameter - a local field $V(x)$ which carries the magnetic
charge - is also readily constructed. It has a form of the singular gauge
transformation operator with the singularity at the point $x$
\begin{equation}
V(x)=\exp {\frac{i}{g}}\int d^2y\
\left[\epsilon_{ij}{\frac{(x-y)_j}{(x-y)^2}}
E_i(y)+\Theta(x-y)J_0(y)\right]  
\label{vqed1}
\end{equation}
where $\Theta(x-y)$ is the polar angle function and $J_0$ is the
electric 
charge density of whatever matter fields
are present in the theory. The cut discontinuity in the function $\Theta$
is not physical and can be chosen parallel to the horizontal axis.
Using the Gauss' law constraint this can be cast in a different form,
which we will find more convenient for our discussion
\begin{equation}
V(x)=\exp {\frac{2\pi i}{g}}\int_C dy^i\
\epsilon_{ij}E_i(y)
\label{vqed}
\end{equation}
where the integration goes along the cut of the function $\Theta$
which starts at the point $x$ and goes to spatial infinity. 
The
operator does not depend on where precisely one chooses the 
cut to lie. 
To see this, note that 
changing the position of the cut $C$ to $C'$ adds to the phase 
${2\pi\over g}\int_{S}d^2x \partial_iE^i$ where $S$ is the area
bounded 
by $C-C'$. 
In the
theory we consider 
only charged particles with charges multiples of $g$ are present. Therefore 
the charge within any closed area is a
multiple integer of the gauge coupling $ \int_{S}d^2x \partial_iE^i=gn$
and the extra phase factor is always unity.

The meaning of the operator $V$ is very simple. From the commutation
relation
\begin{equation}
V(x)B(y)V^\dagger(x)=B(y)+{2\pi\over g}\delta^2(x-y)
\end{equation}
it is obvious that $V$ creates a pointlike magnetic vortex of flux $2\pi/g$.
Despite its nonlocal appearance the operator $V$
can be proven to be a local Lorentz scalar field\cite{kovner2}. The
locality is the consequence of the fact that $V(x)$ commutes with any
local gauge invariant operator in the theory $O(y)$ except when
$x=y$. This is due to the coefficient $2\pi/g$ 
in the exponential which ensures that the Aharonov-Bohm phase of the
vortex created by $V$ and any dynamical charged particle present in
the theory vanishes.
Eqs.(\ref{phiqed},\ref{vqed}) formalize the physical arguments of 't Hooft in
the abelian case.

\subsection{The non-Abelian case at weak coupling.}

Let us now move onto the analogous construction for nonabelian theories.
Ultimately we are
interested in the pure Yang - Mills theory. It is however                 
illuminating to start with the theory with an adjoint Higgs field and
take the decoupling limit explicitly later. For simplicity we
discuss the $SU(2)$ gauge theory.
Consider the
Georgi-Glashow model - $SU(2)$ gauge theory with an adjoint Higgs
field.
\begin{equation}
{\cal L}=-\frac{1}{4}F_{\mu \nu }^{a}F^{a\mu \nu }+\frac{1}{2}({\cal D}_{\mu
}^{ab}H^{b})^{2}+\tilde{\mu}^{2}H^{2}-\tilde{\lambda}(H^{2})^{2}  \label{lgg}
\end{equation}
where 
\begin{equation}
{\cal D}_{\mu }^{ab}H^{b}=\partial _{\mu }H^{a}-gf^{abc}A_{\mu }^{b}H^{c}
\end{equation}

At large and positive $\tilde{\mu}^{2}$ the model is weakly coupled. 
The $SU(2)$ gauge symmetry is broken down
to $U(1)$ and the Higgs mechanism takes place. Two gauge bosons, $W^{\pm }$,
acquire a mass, while the third one, the ``photon'', remains massless
to all orders in perturbation theory. 
The theory in this region of parameter space resembles very
much electrodynamics with vector charged fields.
The Abelian construction can therefore be repeated.
The $SU(2)$ gauge invariant analog of the conserved dual field strength is
\begin{equation}
\tilde{F}^{\mu }={1\over 2}[\epsilon_{\mu\nu\lambda}F^a_{\nu\lambda}n^a-
\frac{1}{g}\epsilon ^{\mu \nu \lambda}
\epsilon ^{abc}n_{a}({\cal D}_{\nu }n)^{b}({\cal D}_{\lambda }n)^{c} ]
 \label{F}
\end{equation}
where $n^a\equiv{\frac{H^a}{|H|}}$ is the unit vector in the
direction of the Higgs field.
Classically
this current satisfies the conservation equation
\begin{equation}
\partial^\mu\tilde F_\mu=0
\label{f}
\end{equation}
The easiest way to see this is to choose a unitary gauge of the form 
$H^a(x)=H(x)\delta^{a3}$. In this gauge $\tilde F$ is equal to the
abelian part of the dual field strength in the third direction in
colour space. Its conservation then follows by the Bianchi
identity. 
Thus classically the theory has a conserved $U(1)$ magnetic charge
$\Phi=\int d^2x \tilde F_0$ just like QED.
However the unitary gauge can not be imposed at the points where $H$
vanishes,
which necessarilly happens in the core of an 't Hooft-Polyakov
monopole. It is well known of course \cite{polyakov} that the
monopoles are the
most important nonperturbative configurations in this model. Their presence
leads to a nonvanishing small mass for the photon
as well as
to confinement of the charged gauge bosons with a tiny
nonperturbative string tension.
As far as the monopole effects on the magnetic flux, 
their presence leads to a
quantum anomaly in the conservation equation (\ref{f}). As a result only
the discrete $Z_2$ subgroup
of the transformation group generated by $\Phi$
remains
unbroken in the quantum theory. The detailed discussion of this
anomaly, the residual $Z_2$ symmetry  and their 
relation to monopoles is given in \cite{kovner1}.

The order parameter for the magnetic $Z_2$ symmetry is constructed
analogously to QED as a singular gauge transformation generated by the
 gauge invariant electric charge operator
\begin{equation}
J^\mu=\epsilon^{\mu\nu\lambda}\partial_\nu (\tilde F^a_{\lambda}n^a), 
\ \ \ \ Q=\int d^2x J_0(x)  
\label{qqcd}
\end{equation}
Explicitly
\begin{eqnarray}
\nonumber
V(x)&=&\exp {\frac{i}{g}}\int d^2y\ \left[\epsilon_{ij}{\frac{(x-y)_j}{(x-y)^2}
} n^a(y)E^a_i(y)+\Theta(x-y)J_0(y)\right]  \\
&=&\exp {\frac{2\pi i}{g}}\int_C dy^i
\epsilon_{ij}n^aE^a_j(y)
\label{VQCD}
\end{eqnarray}
One can think of it as 
a singular $SU(2)$ gauge transformation with the field
dependent gauge function 
\begin{equation}
\lambda^a(y)={\frac{1}{g}}\Theta(x-y)n^a(\vec y)  
\label{lambda}
\end{equation}
This field dependence of the gauge function ensures the gauge 
invariance of the operator $V$.
Just like in QED it can be shown\cite{kovner1}, \cite{kovner2} that
the operator $V$ is a local scalar field.
Again like in QED, the vortex operator $V$ is a local eigenoperator of
the abelian magnetic field $B(x)=\tilde{F}_{0}$. 
\begin{equation}
\lbrack V(x),B(y)]=-{\frac{2\pi }{g}}V(x)\delta ^{2}(x-y)  \label{com}
\end{equation}
That is to say, when acting on a state it creates a pointlike magnetic vortex
which carries a quantized unit of magnetic flux. The $Z_2$ magnetic
symmetry transformation is generated by the operator 
\begin{equation}
U=\exp\{i{g\over 2}\Phi\}
\end{equation}
and acts
on the vortex field $V$ as a phase rotation by $\pi$
\begin{equation}
e^{i{g\over 2} \Phi }V(x)e^{-i{g\over 2} \Phi }=-V(x)
\label{magntr}
\end{equation}

An operator closely related to $U$ and
which will be of interest to us in the following, is
the generator of the magnetic $Z_2$ transformation only inside
some closed contour $C$
\begin{equation}
U(C)=\exp\{i{g\over 2}\int_Sd^2xB(x)\}
\label{uc}
\end{equation}
where the integration is over the area $S$ bounded by $C$. The analog
of 
the commutator eq.(\ref{magntr})
 for this operator is
\begin{eqnarray}
U_CV(x)U^\dagger_C=-&V(x)&,\ \  x\in S\nonumber \\
&V(x)&, \ \ x\notin S
\end{eqnarray}
Taking the contour $C$ to run at infinity $U_C$ becomes the generator of $Z_2$.

We now have the explicit realization of the magnetic $Z_2$ symmetry in 
the Georgi-Glashow model. 

\subsection{The pure gauge theory.}

Our next step is to move on to the pure Yang
Mills theory.
This is achieved by smoothly varying the 
$\tilde\mu^2$ coefficient in the Lagrangian so that the coeficient of
the
mass term of the 
Higgs field becomes positive and eventually arbitrarily large.
It is well known that
in this model
the weakly coupled Higgs regime and strongly coupled
confining regime are not separated by a phase 
transition\cite{fradkin}. 
The pure Yang Mills limit in this model is therefore smooth.

In the pure Yang Mills limit the expressions eq.(\ref{F},\ref{VQCD},\ref{uc})
have to be taken with care. 
When the mass of the Higgs field is very large, the 
configurations 
that dominate the path integral are those with very small value of the
modulus of
the Higgs field $|H|\propto 1/M$.
The modulus of the Higgs field in turn controls the fluctuations of
the unit vector $n^a$,
since the kinetic term for $n$ in the Lagrangian is $|H|^2(D_\mu n)^2$.
Thus as the mass of the Higgs field increases the fluctuations of $n$
grow in both, amplitude and frequency and the magnetic field operator $B$ as defined in
eq.(\ref{F}) fluctuates wildly.
This situation is of course not unusual. It happens whenever one
wants to consider in the effective low energy theory an operator which
explicitly depends on fast, high energy variables. The standard way to deal
with it is to integrate over the fast variables. 
There could be two possible outcomes of this procedure. Either the
operator in question becomes trivial (if it depends strongly on the
fast variables), or its reduced version is well defined and regular on
the low energy Hilbert space.
The ``magnetic field'' operator $B$ in eq.(\ref{F}) is obviously of
the first type. Since in the pure Yang Mills limit all 
the orientations of $n^a$ are equally probable, integrating over the
Higgs field at fixed $A_\mu$ will lead to vanishing of $B$. However
what interests us is not so much the magnetic field but rather the
generator of the magnetic $Z_2$ transformation $U_C$ of
eq.(\ref{uc}). In the pure Yang-Mills limit we are thus lead to
consider the operator
\begin{eqnarray}
U_C=lim_{H\rightarrow 0}\int Dn^a&&\exp\Big\{-|H|^2(\vec D n_a)^2\Big\}\\
&&\exp\Big\{i{g\over 4}\int_C d^2x\big(\epsilon_{ij}F^a_{ij}n^a-
\frac{1}{g}\epsilon ^{ij}\epsilon ^{abc}n_{a}({\cal D}_{i }n)^{b}
({\cal D}_{j }n)^{c}\big ) \Big\}\nonumber
\label{uc1}
\end{eqnarray}
The weight for the integration over $n$ is the kinetic term for the isovector $n_a$. As was noted
before the action does not depend on $n^a$ in the YM limit.
This term however regulates the integral and we keep it for this reason.
This operator
may look somewhat unfamiliar at first sight. However in a remarkable
paper \cite{dp} Diakonov and Petrov showed that eq.(\ref{uc1}) is equal to 
the trace of the fundamental Wilson loop along the contour 
$C$\footnote{We note that Dyakonov and Petrov had to introduce a regulator
to define the path integral over $n$. The regulator they required
was precisely of the same form as in eq.(\ref{uc1}). 
It is pleasing to see that this regulator appears
 naturally in our approach as the remnant of the kinetic term of the 
Higgs field.}.
\begin{equation}
U_C=W_C\equiv\Tr{\cal P}\exp \Big\{ig\int_C dl^iA^i\Big\}
\end{equation}

We conclude, that in the pure Yang-Mills theory the generator of the magnetic 
$Z_2$ symmetry is the fundamental spatial Wilson loop along the boundary of 
the spatial plane.

There is a slight subtlety here that may be worth mentioning.
The generator of a unitary transformation should be a unitary operator. The trace of the 
fundamental Wilson loop on the other hand is not unitary. One should therefore
strictly speaking consider instead a
unitarized Wilson loop $\tilde W={W\over \sqrt{WW^\dagger}}$. However the factor
between the two operators $\sqrt {WW^\dagger}$ is an operator that is only sensitive
to behaviour of the fields at infinity. It commutes with all physical local operators
$O(x)$ unless $x\rightarrow\infty$. 
In this it is very different from the Wilson loop itself,
which has a nontrivial commutator with vortex operators $V(x)$ at all values of $x$.
Since the correlators of all gauge invariant local fields in the pure Yang Mills theory
are massive and therefore short range, the operator $\sqrt{WW^\dagger}$ must be a constant
operator on all finite energy states. The difference between $W$ and $\tilde W$ is 
therefore a trivial constant factor and we will not bother with it in the following.
Perhaps of more concern is the difference between $W_C$ and $\tilde W_C$ when the contour
$C$ is not at infinity. However here
again the factor between the two operators $\sqrt {W_CW_C^\dagger}$ 
is only sensitive to physical degrees of freedom
on the contour $C$ and not inside it. Due to its presence the vacuum averages 
of $W_C$ and $\tilde W_C$ may
differ at most by a factor which has a perimeter behaviour 
$<W_C>=\exp\{mP(C)\}<\tilde W_C>$
where $P(C)$ is a perimeter of $C$. The question we will be interested in
is whether $<W_C>$  has a perimeter or area behaviour. As far as the answer to this 
question is concerned $W_C$ and $\tilde W_C$ are completely equivalent, and we will not
make distinction between them. In the rest of this paper we will therefore refer to $W$
as the generator of $Z_2$ keeping this little caveat in mind.

Next we consider the vortex operator eq.(\ref{VQCD}). Again we have to integrate 
it over 
the orientations of the unit vector $n^a$. 
This integration in fact is equivalent to averaging over the gauge group.
Following \cite{dp} one can write  $n_a$ in terms of the SU(2)
gauge transformation matrix $\Omega$.
\begin{equation}
\vec n={1\over 2}{\rm Tr}\Omega\tau\Omega^{\dagger}\tau_3
\label{ucl3}
\end{equation}
The vortex operator in the pure gluodynamics limit then becomes
\begin{equation}
\tilde V(x)=\int D\Omega\exp{{2\pi i\over g}\int_C
dy_i\epsilon_{ij}{\rm Tr}\Omega E_j\Omega^{\dagger}\tau_3}
\end{equation}
This form makes it explicit that $\tilde V(x)$ is defined as the gauge singlet part
of the following, apparently non gauge invariant operator
\begin{equation}
V(x)=\exp {\frac{2\pi i}{g}}\int_C dy^i
\epsilon_{ij}E^3_i(y)
\label{v33}
\end{equation}
The integration over $\Omega$ obviously projects out the gauge singlet part of $V$.
In the present case however this projection is redundant. This is because
even though $V$ itself is not gauge invariant,
when acting on 
a physical state it transforms it into another physical 
state\footnote{This is not a trivial statement, since a generic 
nongauge invariant operator has 
nonvanishing matrix elements between the physical and an unphysical sectors.}. 
By physical states
we mean the states which 
satisfy the Gauss' constraint in the pure Yang-Mills theory.
This property of $V$ was noticed by 't Hooft \cite{thooft}.
To show this let us consider $V(x)$ as defined 
in eq.(\ref{v33})
and its gauge transform $V_\Omega=\Omega^\dagger V \Omega$ where $\Omega$ is an arbitrary
nonsingular gauge transformation operator.
The wave functional of any physical state depends only on gauge invariant characteristics
of the vector potential, i.e. only on the values of Wilson loops over all possible 
contours. 
\begin{equation}
\Psi[A_i]=\Psi[\{W(C)\}]
\end{equation}
Acting on this state by the operators $V$ and $V_\Omega$ respectively we obtain
\begin{eqnarray}
V|\Psi>&=&\Psi_V[A_i]=\Psi[\{VW(C)V^\dagger\}] \nonumber \\
V_\Omega|\Psi>&=&\Psi_V^\Omega[A_i]=\Psi[\{V_\Omega W(C)V^\dagger_\Omega\}]
\end{eqnarray}
It is however easy to see that the action of $V(x)$ and $V_\Omega(x)$ on 
the Wilson loop is 
identical - they both multiply it by the centergroup phase  (which stays unaffected by $\Omega$) if $x$ is inside $C$ and do 
nothing otherwise.
Therefore we see that
\begin{equation}
V|\Psi>=V_\Omega|\Psi>
\label{prev}
\end{equation} 
for any physical state $\Psi$.
Thus we have
\begin{equation}
\Omega V|\Psi>= \Omega V\Omega^\dagger|\Psi>=
V|\Psi>
\end{equation}
where the first equality follows from the fact that a physical state is invariant under
action of any gauge transformation $\Omega$ and the second equality 
follows from eq.(\ref{prev}).
But this equation is nothing but the statement that the state
$V|\Psi>$ is physical, i.e. invariant under any nonsingular gauge
transformation. 

We have therefore proved that 
when acting on a physical state the vortex operator creates 
another physical state. 
For an operator of this type the gauge invariant projection only affects
its matrix elements between unphysical states. Since we are only
interested
in calculating correlators of $V$ between physical states, the
gauge projection is redundant and we can freely use $V$ rather than $\tilde V$ to represent the vortex operator.

It is instructive to note that this propery is not shared by the Wilson loop. One can 
in fact represent the Wilson loop as a singlet gauge projection of a simple Abelian  
loop operator. The second exponential in eq.(\ref{uc1}) can be written as
\begin{equation}
\exp\Big\{i{g\over 2}\int_C dl^i A^i_an^a-{i\over 2}\int d^2x\epsilon_{ij}\epsilon^{abc}n_a\partial_i n_b\partial_j n_c\Big\} 
\label{ucl2}
\end{equation} 
Using eq.(\ref{ucl3}) 
we can rewrite the integral in eq.(\ref{uc1}) -omitting the regulating kinetic piece-
as:
\begin{equation}
W_C=\int D\Omega\exp\Big\{i{g\over 2}\int {\rm Tr}\tau_3\big( \Omega A^i\Omega^{\dagger}+i\Omega\partial^i\Omega^{\dagger}\big)dl^i\Big\}
\end{equation}

The Wilson loop is therefore the gauge singlet part of the Abelian loop
\begin{equation}
U^A_C=\exp{i{g\over 2}\int {\rm Tr}A^i\tau_3dl^i}
\end{equation}
The matrix elements of $W_C$ and $U^A_C$ on physical subspace therefore are the same.
However $U^A_C$ 
as opposed to $V$ does have nonvanishing nondiagonal 
matrix elements, that is matrix elements between
the physical and the unphysical sectors. It therefore can {\it{not}} be used 
instead of $W_C$ in gauge theory calculations. For example non
gauge invariant states will contribute as intermediate states in the 
calculation of
quantities like the correlation function $<U^A_{C1}U^A_{C_2}>$,
while their contribution vanishes in similar correlators which involve the Wilson loop.

The generalization of the preceding discussion to $SU(N)$ gauge theories 
is straightforward. Once again one can start with the Georgi-Glashow like model, 
where  the $SU(N)$ is higgsed to $U(1)^{(N-1)}$\footnote{In $SU(N)$ 
theories with $N>2$
there in principle 
can be phases separated from each other
due to spontaneous breaking of some global symmetries.
For instance the $SU(3)$ gauge theory with adjoint matter has
a phase with spontaneously broken
charge conjugation invariance
\cite{bronoff}. Still even in this phase
the confining properties  are the same as in 
the strongly coupled pure Yang-Mills theory,
with the Wilson loop having an area law.}. The construction of the vortex
operator and the generator of $Z_N$ in this case is very similar and the details are
given in \cite{kovner1}.
Taking the mass of the Higgs field to infinity again projects the generator onto the 
trace of the fundamental Wilson loop. The vortex operator can be taken as 
\begin{equation}
V(x)=\exp\{{ 4\pi i\over gN} \int_C dy^i
\epsilon_{ij}\Tr(YE_i(y))
\label{v2}
\end{equation}
where the hypercharge generator $Y$ is defined as
\begin{equation}
Y={\rm diag} \left(1,1,...,-(N-1)\right)
\end{equation}
and the electric field is taken in the matrix notation $E_i=\lambda^aE^a_i$ with
$\lambda^a$ - the $SU(N)$ generator matrices in the fundamental representation.

\subsection{Generalization to 3+1 dimensions}

To conclude this section we discuss how the magnetic symmetry 
structure generalizes to four dimensions.
The conserved $Z_N$ generator  in the Georgi-Glashow model is defined through
\begin{equation}
U_S=\exp\Big\{i{g\over 2}\int_S d^2S^i\big (B^a_{i}n^a-
\frac{1}{g}\epsilon ^{ijk}
\epsilon ^{abc}n_{a}({\cal D}_{j}n)^{b}({\cal D}_{k }n)^{c}\big ) \Big\}
\end{equation}
Although the definition of $U$ contains explicitly the surface $S$ through which the 
abelian magnetic flux is integrated, the operator in fact does not depend on $S$
but is specified completely by its boundary. This is because changing $S$ changes the 
phase 
of $U$ by the magnetic flux through the closed surface. The only dynamical objects that
carry magnetic flux in the theory are 't Hooft-Polyakov monopoles. Since their flux is 
quantized
in units of $4\pi/ g$ the change in the phase is always a multiple integer of $2\pi$.
In the pure Yang-Mills limit the operator $U_S$ again reduces to the trace of the 
fundamental
Wilson loop along the boundary of $S$.
Taking the contour to infinity defines the generator of magnetic $Z_N$. As we have already 
noted, this charge is a little unusual in that it is defined as a surface integral 
rather than a volume integral. As a result the order parameter
for this symmetry transformation is not a local but rather a stringy field.
This is of course just a restatement of the fact that magnetic vortices in 3+1 dimensions
are stringlike objects. 
The operator that creates a vortex can still be defined in a way similar to 2+1 dimensions.
Skipping the intermediate steps which we went through in the previous discussion
we give the final result for the pure Yang Mills $SU(N)$ gauge theory. The magnetic vortex
along the curve $C$ is created by the following operator of the 
"singular gauge transformation"\footnote{The derivative term $\partial^i\omega$ 
in this expression should be understood to contain only the smooth part of the derivative
and to exclude the contribution due to the discontinuity of $\omega$ on the surface $S$.}
\begin{equation}
V(C)=\exp\{{ i\over gN} \int d^3x \Tr(D^i\omega_CY) E^i\}=
\exp\{{4\pi i\over gN}\int_S d^2 S^i \Tr (YE^i)\}
\label{v3}
\end{equation}
with $\omega_C(x)$, the singular gauge function 
which is equal to the solid angle subtended by $C$ as seen from the 
point $x$. The function $\omega$ is continuos everywhere, except on a surface $S$
bounded by $C$, where it jumps by $4\pi$. Other than the 
fact that $S$ is bounded by 
$C$, its location is arbitrary. 
The vortex loop and the spatial Wilson loop satisfy the 't Hooft algebra
\begin{equation}
V^\dagger(C)W(C')V(C)=e^{{2\pi i\over N} n(C,C')}W(C')
\end{equation}
where $n(C,C')$ is the linking number of the curves $C$ and $C'$.
One can consider closed contours $C$ or infinite contours that run through 
the whole system. 
For an infinite contour $C$ and the Wilson loop along the spatial
boundary of the system
the linking number is always unity.
The $V(C)$ for an infinite loop is therefore an eigenoperator
of the $Z_N$ magnetic symmetry and is the analog of the vortex
operator 
$V(x)$
in 2+1 dimensions. Any closed vortex loop of fixed size commutes with the 
Wilson loop if the contour $C'$ is very large. Such a closed loop is thus
an  analog of the vortex-antivortex correlator
$V(x)V^\dagger(y)$, which also commutes with the global symmetry
generator, 
but has a
nontrivial
commutator with $U_C$ if $C$ encloses only one of the points $x$ or $y$.

To summarize this section, we have shown that  pure Yang Mills 
theory in 2+1 and 3+1 
dimensions
has a global $Z_N$ magnetic symmetry. The generator of the symmetry 
group in both cases
is the trace of the fundamental Wilson loop along the spatial boundary
of the system.
The order parameter for this symmetry in 2+1 dimensions is a local
scalar 
field $V(x)$,
while in 3+1 dimensions a stringlike field $V(C)$. In both cases the
field 
$V$ is gauge 
invariant on physical states and is a {\it bona fide} canonical order
 parameter
which distinguish in gauge invariant way the phases of the theory.
In the next section we discuss 
the realization of the magnetic symmetry in the confined
and the deconfined phases.

\section{Hot and cold realizations of the magnetic $Z_N$.}

As with any global symmetry, it is important to understand what is the
mode of realization of magnetic $Z_N$ in the ground state of the
theory. This mode of realization depends of course on the parameters
of the theory as well as on the temperature.
The situation at zero temperature is well understood.

\subsection{2+1 dimensions.}

Again we start with three dimensions. 
There is a very general argument due to
't Hooft\cite{thooft}\footnote{The original argument as stated in
\cite{thooft} is formulated for 3+1 dimensional theories, however its 
generalization to 2+1 dimensions requires only linguistic changes.} 
stating
that if the theory does not have zero mass excitations
the area law of the  Wilson loop implies the nonvanishing expectation
value of the vortex operator $V(x)$. Conversely if the Wilson loop has
a perimeter law the expectation value of $V(x)$ must vanish and the
correlation function $V(x)V^\dagger(y)$ must have an exponential
falloff with $|x-y|$.
It follows that in the pure Yang Mills theory the vacuum
expectation value of the vortex operator does not vanish and therefore
the $Z_N$ magnetic symmetry is spontaneously broken. The same is true
in the partially broken Higgs phase of the Georgi-Glashow model. As
mentioned in the last section the confining and the Higgs regimes in
this model are analytically connected and therefore the realization of
all global symmetries in the two regimes is the same. 

In fact in the
weakly coupled Higgs phase this can be verified by the direct
calculation of the expectation value of $V$ \cite{kovner1}. This
calculation maps very simply into the classic monopole plasma
calculation of Polyakov and was discussed in detail in
\cite{kovner1}. One can also explicitly construct the low energy
effective Lagrangian in terms of the field $V$ which realizes the
spontaneously broken $Z_N$ symmetry and describes the low energy
spectrum of the Georgi - Glashow vacuum.
\begin{equation}
{\cal L}=\partial_\mu V^*\partial^\mu V -
\lambda(V^*V-\mu^2)^2 -\zeta (V^2 +V^{*2}) 
\label{efl}
\end{equation}
A similar effective Lagrangian
with some quantitative differences was argued to be valid
also for the pure Yang-Mills theory in \cite{kovner3}.

The application of the 't Hooft argument at finite temperature is somewhat less 
straightforward. 
Since
at finite temperature the Lorentz invariance is broken, the temporal
and
spatial Wilson loops do not necessarily have the same behaviour and
one has to be more careful. The original argument relates the behaviour
of the vortex
operator and the temporal Wilson loop. At finite temperature in the Euclidean 
formalism the
extent of the system in the temporal direction is finite. As a result
it is not possible to distinguish between the area and perimeter law 
for "asymptotically" 
large temporal loops. Instead the role of the temporal Wilson loop is 
taken over by the
Polyakov line - the loop that winds around the total volume of 
the system in the temporal 
direction.
Thus one expects that in the
deconfining phase where the Polyakov line has a nonvanishing vacuum average, 
the vortex operator should have vanishing expectation
value. Indeed this can be easily confirmed by the 
explicit calculation of the VEV of the vortex
operator using the method of \cite{KAKS}. 
In \cite{KAKS} the calculation was performed in 3+1 dimensions, but
adapting it to 2+1 dimensional case is trivial. We give below a brief outline.

Consider the equal time vortex-antivortex correlation function.
At finite temperature is it given by the 
following expression
\begin{equation}
<V(x)V^\dagger(y)>={\rm Tr} e^{-{\beta\over 2}(E^2+B^2) }
e^{i{2\pi i\over g}\int_x^y\epsilon_{ij} dl^i E^j_3}
\label{vt}
\end{equation}
The line integral can be taken along the straight line $L$  connecting the
points $x$ and $y$. For definiteness we take $x$ and $y$ to be
separated in the direction of the first axis.
Introducing the imaginary time axis and the 
Lagrange multiplier field $A_0$ in the standard way
this expression can be transformed to
\begin{equation}
<V(x)V^\dagger(y)>
=\int DA_iDA_0\exp\{-{1\over 2}\int_0^{\beta} dt\int d^3x
\left(\partial_0A_i^a-(D_iA_0)^a-\delta(t)a_i^a\right)^2+(B^a)^2\}
\label{vt1}
\end{equation}
where the ``external field `` $a^i$ is given by
\begin{equation}
a^a_i({\bf x})=
\delta^{a3}\delta_{i2}{2\pi\over g}\delta({\bf x}-L)
\label{a}
\end{equation}
The delta function in time
in front of the external field $a^i$ in eq.(\ref{vt1}) appears for the
following reason.
As we saw in section 2.3 
the product of the vortex and an antivortex operator is gauge
invariant. This is because it induces a singular gauge
transformation
which is continuous up to the center element.
However if we split it up in imaginary time into infinitesimal  
bits $V_{dt}=e^{idt T{2\pi i\over g}\int_x^y\epsilon_{ij} dl^i E^j_3}$ 
then any single such bit separately is not gauge invariant,
since the
transformation it induces is genuinely discontinuous across the line
connecting the points $x$ and $y$.
The operators $V_{dt}$ therefore do not commute with the projection
operator on physical states. 
To obtain the path integral representation for the
expectation value eq.(\ref{vt}) we should introduce the projection
operator only at the last point in imaginary time and not at the
intermediate points. In the path integral language 
this corresponds to the gauge fixing $A_0=0$ everywhere except at one
time, say $t=0$. In this gauge it is straightforward to see that the
Gaussian integration over the electric field leads to the usual path
integral with $\partial_iA_0$ shifted by $a_i$. This can then be
rotated to an arbitrary gauge with the result
eq.(\ref{vt1})\footnote{In 
this derivation we dropped commutator terms between the Hamiltonian and
the exponent in the vortex-antivortex operator. 
These commutator terms only exist at $t=0$
and therefore drop out in the continuum limit (i.e are $ O(dt)$). In the 
lattice realization they are indeed present and complete the
expression
eq.(\ref{vt1}) to the twisted plaquette 
representation\cite{michels}}.

To  evaluate the path integral eq.(\ref{vt1}) we follow the standard
procedure and integrate out all modes except for the Polyakov
loop in a saddle point approximation\cite{weiss},\cite{korthals}.
This leads to the effective action
$S_{eff}(q,a_i)$, where $1/2TrP=\cos q$
\begin{equation}
<V(x)V^\dagger(y)>=\int Dq\exp -S_{eff}(q,a_i)
\end{equation}
To one loop order the effective action is given by
\begin{equation}
S_{eff}=\int d^2x\big({2T\over g^2}(\partial_iq+{g\over 2}a_i)^2+U(q)\big)
\label{action}
\end{equation}
The matrices 
$\tau^a$ in $A_{\mu}$ are the generators of $SU(2)$ in the 
fundamental representation
and are normalized according to ${\rm tr}\tau^a\tau^b={1\over
2}\delta^{ab}$.

The one loop effective potential $U$ is related to a Bernoulli
polynomial and can be read off the expressions 
in \cite{korthals}~\cite{michels}. The
only property of $U$ which is important to us is that it has two
degenerate minima at $q=0,\pi$.

To calculate the correlator we have to find the configuration of $q$ 
which minimizes the action eq.(\ref{action}).
Qualitatively the form of the solution is clear. The 
considerations identical to those in \cite{KAKS}
tell us that it must be the ``broken'' electric $Z_2$ domain wall:
half a wall ( $q\rightarrow_{x_2\rightarrow\infty}0$,
$q(x_2=0)={\pi\over 2}$) 
above the line $L$ and half a wall
( $q(x_2=0)=-{\pi\over 2}$, $q\rightarrow_{x_2\rightarrow-\infty}0$,) 
below the line $L$ separated by a discontinuity $\delta q=\pi$.
The action of such a configuration is $S_{eff}=\tilde \alpha|x-y|$
where $\tilde\alpha$ is the ``$Z_2$ domain wall tension''. The
vortex correlator is thus given by
\begin{equation}
<V(x)V^\dagger(y)>=\exp\{-\tilde\alpha|x-y|\}
\end{equation}
As $|x-y|$ become large the correlation function decreases
exponentially, and thus the expectation value of the vortex operator
vanishes.
For the $SU(N)$ group this calculation trivially generalizes and gives
the same result.
The exponential decay is also obtained for the correlator of $V^m$ with any power 
$m<N$.

Recall that the vortex operator is the order parameter for the
magnetic $Z_N$ symmetry. Moreover the powers of $V$ exhaust all possible 
local order parameters\footnote{The latter statement is correct 
modulo multiplication of $V^m$ by local gauge invariant and $Z_N$ invariant operators.
These possible factors do not change the fact of the exponential decay
of the corelators and are therefore unimportant for our discussion.}.
Their vanishing is therefore an unambiguous
indication that the magnetic $Z_N$ is restored in the high temperature
deconfined phase.
 
In hindsight this is not very surprising. Indeed, we are dealing with
physical discrete symmetry which is spontaneously broken at zero temperature.
When the system is heated it is unavoidable that entropy effects take
over and at some sufficiently high temperature the symmetry must be
restored.
A good qualitative guide here is the effective Lagrangian
eq.({\ref{efl}).
It describes a simple $Z_2$ invariant scalar theory. There is very
little doubt that a system described by this Lagrangian indeed
undergoes a symmetry restoring phase transition at some $T_c$.
Moreover the effective Lagrangian approach also suggests that this 
phase transition has
deconfining character. As shown in \cite{kovner1,kovner2,kovner3} the 
charged states
in the effective theory eq.(\ref{efl}) are represented by solitonic
configurations 
of the vortex 
field $V$ with unit winding number. The energy of any such state is
linearly 
divergent in 
the infrared. The reason is that due to finite degeneracy of vacuum
states, the minimum
energy configuration looks like a quasi onedimensional strip across
which the 
phase of $V$
winds. The energy density inside this "electric flux tube" is
proportional to the vacuum expectation
value of $V$. When the VEV vanishes, so does the string
tension. Stated in other 
words,
when $V$ vanishes, the phase fluctuations are large and the winding number is
not a sharp observable. An external charge is thus 
screened easily by regions of space around it with vanishing $V$ .
The phase with $<V>=0$ is therefore not confining. In the theory with several Higgs
fields this phase exists even at zero temperature and corresponds to
a completely Higgsed
phase - where the gauge group is broken completely. In such a Higgs phase indeed 
the colour is screened rather than confined. In the pure Yang-Mills
theory this 
phase 
is absent at zero temperature, but is realized as the deconfined 
phase at $T>T_c$.
We thus see that the behaviour of the vortex operator at high
temperature does indeed match the simple intuition coming from a $Z_N$
invariant effective Lagrangian very well.

\subsection{Extension to 3+1 dimensions}

The 't Hooft argument now states that
the vanishing vacuum average of the Polyakov line
is incompatible with the area law
behaviour of the spatial 't Hooft loop and vice versa. 
This means that in the confining phase the 't Hooft loop has
perimeter law. In the high temperature deconfined phase the behaviour
of the spatial 't Hooft
loop must become area since the average of the Polyakov line is finite.
Again this is confirmed by explicit calculation in \cite{KAKS}.

A more subtle question is how the behaviour of the 't Hooft
loop relates to the realization of the magnetic $Z_N$ symmetry. The
$Z_N$ symmetry does not have an order parameter which is a local field
defined at a point. The only order parameters in the strict sense
(an eigenoperator with a nonvanishing eigenvalue) is a 't Hooft
line $V(C)$ which runs through the whole system~\cite{thooftperiodic}. 

In a 
system which is finite in the direction of the loop, but is infinite
in the perpendicular directions everything is clear cut. 
In this case there are two
possibilities: 

a) $<V>\ne 0$  and the magnetic $Z_N$ broken, or

b). $<V>=0$ and the nagnetic $Z_N$ restored.

In the system infinite in all directions  $C$ is necessarily 
an infinite line, and 
the expectation value $<V(C)>$
clearly vanishes irrespective of whether $Z_N$ is broken or not.
The 't Hooft loop along a closed contour on the other hand is never
zero, since it is globally invariant under the $Z_N$ transformation. 
It is therfore impossible to find an operator whose VEV distinguishes
between the two possible realizations of the magnetic
symmetry by vanishing in one phase and not vanishing in the other.
Nevertheless the behaviour of the closed loop does
indeed reflect the realization mode of the symmetry, since it is
qualitatively different in the two possible phases. 
Namely vacuum expectation value of a large closed 't Hooft loop (by large, as usual we mean  
that the linear dimensions of the loop are much larger than the
correlation length in the theory) has an area law decay if the
magnetic symmetry is spontaneously broken, and perimeter law decay if
the vacuum state is invariant.

To understand the physics of this 
behaviour
it is useful to think of the 't Hooft line
as built of ``local'' operators - little ``magnetic dipoles''.
Consider eq.(\ref{v3}) with the contour $C$ running
along
the x- axis and the surface $S$ chosen as the $(x,y)$ plane. 
Let us mentally divide the line into (short) segments of length $2\Delta$
centered at $x_i$. Each one of these segments is a little magnetic
dipole and the 't Hooft loop is a product of the operators that
create these dipoles. The definition of these little dipole operators 
is somewhat ambiguous but since we only intend to use them
here
for the purpose of an intuitive argument any reasonable definition will do.
It is convenient to define a single dipole operator in the following way
\begin{equation}
D_\Delta(x)=
\exp
\{i\int d^3y [a_i^+(x+\Delta-y)+a_i^-(x-\Delta-y)]
\Tr (YE^i(y))\}
\label{d}
\end{equation}
where $a_i^\pm(x-y)$ is the c-number vector potential of the 
abelian magnetic monopole (antimonopole) of strength $4\pi/gN$. 
The monopole field correpsonding to $a_i$ contains both, the 
smooth $x_i/x^3$ part as well as the Dirac string contribution. 
The Dirac string
of the monopole - antimonopole pair in 
eq.(\ref{d}) is chosen so that is connects the points $x-\Delta$ and
 $x+\Delta$ along the straight line.
The dipole operators obviously have the property
\begin{equation}
D_\Delta(x)
D_\Delta(x+2\Delta)
=D_{2\Delta}(x+\Delta)
\end{equation}
This is because in the product the smooth field contribution of the monopole in
$D_\Delta(x)$ cancels the antimonopole contribution in 
$D_\Delta(x+2\Delta)$, while the Dirac string now stretches between the points
 $(x-\Delta)$ and $(x+3\Delta)$. When multiplied over the closed contour, 
the smooth fields cancel out completely, while the surviving Dirac string 
is precisely the magnetic vortex created by a closed 't Hooft loop operator. 
The 't Hooft loop can therefore be written as
\begin{equation}
V(C)=\Pi_{x_i}D_\Delta(x_i)
\label{vd}
\end{equation}

The dipole operator $D(x_i)$ is an eigenoperator of the magnetic flux
defined on a surface that crosses the segment $[x_i-\Delta,
x_i+\Delta]$.
 Suppose the magnetic symmetry is broken. Then we expect the dipole operator
to have a nonvanishing expectation value\footnote{The magnetic dipole 
operators defined above are strictly speaking not local, since they carry
the long range magnetic field of a dipole. However, the dipole 
field falls off with distance very fast. Therefore even though this fall off
is not exponential the slight nonlocality of $D$ should not affect the 
following qualitative discussion.}
 $<D>=d(\Delta)$. If there are no
massless excitations in the theory, the operators $D(x_i)$ and
$D(x_j)$ should be decorrelated if the distance $x_i-x_j$ is greater
than the correlation length $l$. Due to eq.(\ref{vd}), the 
expectation value of the 't Hooft loop
should therefore roughly behave as
\begin{equation}
<V(C)>=d(l)^{L/l}=\exp\{-\ln\Big({1\over d(l)}\Big) {L\over l}\}
\label{vline}
\end{equation} 
where $L$ is the perimeter of the loop.
In the system of finite length $L_x$, 
the vacuum expectation value of the vortex line which winds around the
system in $x$-direction is therefore finite as in eq.(\ref{vline})
with $L\rightarrow L_x$.

On the other hand in the unbroken phase 
the VEV of the dipole operator depends on the size of the system in
the perpendicular plane $L_y$. For large $L_y$ it must vanish
exponentially as
$d=\exp\{-a L_y\}$. 
So the expectation
value of $V$  behaves at finite $L_y$ in the unbroken phase as:
\begin{equation}
<V(C)>=\exp\{-a L_yL_x\}
\label{vlinebehave}
\end{equation}
and vanishes as $L_y\rightarrow\infty$.
Thus in a system which is finite in $x$ direction,
but infinite in $y$ direction, the 't Hooft line in the $x$ direction
has a finite VEV in the broken phase and vanishing VEV in the
unbroken phase.

In the limit of the infinite system size $L_x\rightarrow\infty$
the VEV obviously vanishes in both phases. 
This is of course due to the fact that $V$ is a product of infinite
number
of dipole operators, and this product vanishes even if individual
dipole operators have finite VEV\footnote{The VEV of
the dipole $D$ must be smaller than one since $D$ is defined as a
unitary operator.}.
However one can avoid any reference to finite size system
and infinite vortex lines by considering closed 't Hooft loops.

For a closed loop with long sides along $x$ axis at $y=0$ and $y=R$
the above argument leads to the conclusion  
that in the broken phase $V$ must have a
perimeter law, eq.(\ref{vline}). 
In the unbroken phase the correlation between the
dipoles at $y=0$ and dipoles at $y=R$ should decay exponentially
$<D(0)D(R)>\propto\exp\{-\alpha{R\over l}\}$
and thus 
\begin{equation}
<V(C)>=\exp\{-\alpha {LR\over l^2}\}=\exp\{-\alpha {S \over l^2}\}
\end{equation}
Thus the perimeter behaviour of $<V(C)>$ indicates a vacuum state which
breaks spontaneously the magnetic $Z_N$ while the area behaviour means
that the magnetic $Z_N$ is unbroken.

The results of \cite{KAKS} then mean that in 3+1 dimensions as well as
in 2+1 dimension
the magnetic symmetry is restored above the
deconfining phase transition, in the sense of eq.(~\ref{vlinebehave}).

In the next section we discuss what is the implication of this conclusion
on the behaviour of the spatial Wilson loop.

\section{Spatial Wilson loop at low and high temperature.}

As we have shown in Section 2 the spatial Wilson loop is the generator
of the magnetic $Z_N$ symmetry. We expect therefore that the mode of
realization of the magnetic $Z_N$ is strongly linked to
the behaviour of $W$. 
The argument is simplest to state for a 
toy model which exemplifies the basic physics in a very simple setting.

Rather than talk
about nonabelian gauge theory, consider a scalar theory of a complex
field $\phi$ with global
$Z_N$ theory in 2+1 dimensions.
\begin{equation}
{\cal L}=\partial_\mu\phi\partial_\mu\phi^*+\lambda(\phi^*\phi-\mu^2)^2+
\zeta\Big(\phi^N+(\phi^*)^N\Big)
\label{lagr}
\end{equation}
The generator of the $Z_N$ symmetry is given by
\begin{equation}
U=\exp\Big\{i{2\pi\over N}\int d^2xj_0(x)\Big\}=
\exp\Big\{{2\pi\over N}\int d^2x (\pi\phi-\pi^*\phi^*)\Big\}
\end{equation}
where $\pi=\partial_0\phi^*$ is the momentum conjugate to the field
$\phi$.
Obviously with the canonical commutation relations between $\pi$ and
$\phi$ one has
\begin{equation}
U\phi(x)U^\dagger=e^{i{2\pi\over N}}\phi(x)
\end{equation}
We will be interested in the behaviour of the operator which generates
the $Z_N$ transformation only inside some region $S$ of the two
dimensional plane.
\begin{eqnarray}
\label{uloop}
U(S)&=&\exp\Big\{{2\pi\over N}\int_S d^2x (\pi\phi-\pi^*\phi^*)\Big\}\\
U(S)\phi(x)U^\dagger(S)&=&e^{i{2\pi\over N}}\phi(x)\ \  x\in S\nonumber \\
&=&\phi(x), \ \ x\notin S\nonumber
\end{eqnarray}
We will refer to this operator as the U-loop.
Throughout this discussion we assume that there are no massless
excitations in the spectrum of the theory and that the linear
dimensions of the area $S$ are much larger than the correlation length.

The statement we are aiming at 
is that at zero temperature in the phase with broken $Z_N$ the U-loop has an
area law behaviour while in the phase with unbroken $Z_N$ this changes
into the perimeter law behaviour. 

\subsection{U-loop in the broken phase}

Consider the broken phase first.
We are interested in the vacuum expectation value
of $U(S)$. This is nothing but the overlap of the vacuum state $<0|$
and the state which is obtained by acting with $U(S)$ on the vacuum
state $|S>=U|0>$. If the symmetry is broken,
the field
$\phi$ in the vacuum state  is pointing in some fixed direction in the
internal space. In the state $|S>$ on the other hand its direction in
the internal space is different - rotated by $2\pi/N$ - at points
inside the area $S$.
In the local theory with finite correlation length the overlap
between the two states approximatelly factorises into the product of the
overlaps taken over the region of space of linear dimension of order
of the correlation length $l$
\begin{equation}
<0|S>=\Pi_x<0_x|S_x>
\label{fact}
\end{equation}
where the label $x$
is the coordinate of the
point in the center of a given small region of space. For $x$
outside the area $S$ the two states $|0_x>$ and $|S_x>$ are identical and
therefore the overlap is unity. However for $x$ inside $S$ the states
are different and the overlap is therefore some number $e^{-\gamma}$ 
smaller than unity. The number of such regions inside the area is obviously 
of order $S/l^2$
and we thus
\begin{equation}
<U(S)>=\exp\{-\gamma{S\over l^2}\}
\end{equation}
In a weakly coupled theory this argument is confirmed
by explicit calculation. The expectation value of the U-loop in the
theory eq.(\ref{lagr}) is given
by the following path integral
\begin{equation}
<U(C)>=\int d\phi d\phi^*\exp\Bigg\{-\int (\partial_\mu
\phi+i\phi\chi_\mu)(\partial_\mu\phi^*-i\phi^*\chi_\mu)+
\lambda(\phi^*\phi-\mu^2)^2+
\zeta\Big(\phi^N+(\phi^*)^N\Big)\Bigg\}
\label{uexp}
\end{equation}
with
\begin{eqnarray}
\chi_\mu(x)&=&{2\pi\over N}\delta^{\mu 0}\delta(x_0),\ \ \ x\in S \\
&=& 0, \ \ \ x\notin S
\end{eqnarray}
This expression directly follows from eq.(\ref{uloop}) and integration
over the canonical momentum in the phase space path integral.
At weak coupling this path integral is
dominated by a simple classical configuration. First, it is clear that
the solution must be such that the phase of the field $\phi$
has a discontinuity of $2\pi/N$ when crossing the surface
$S$ since otherwise the action is UV divergent due to singular
$\chi$. Asymtotically at large distance from the surface
the field should approach its vacuum expectation value. Since the
source term $\chi$ vanishes outside $S$, eveywhere where $\phi$ is
continuous it has to solve classical equations of
motion. Also, for values of $x_1$ and $x_2$ which are well inside $S$
the profile $\phi$ should not depend on these coordinates, but should
only depend on
$x_0$.
It is easy
to see that a solution with these properties exists:
it is given by the
``broken'' domain wall solution.
Recall that the vacuum is degenerate and so there certainly exists 
a classical solution
of the equations of motion which interpolates between two adjacent
vacuum states $\phi\rightarrow_{x_0\rightarrow\infty}\phi_0$ and 
$\phi\rightarrow_{x_0\rightarrow-\infty}e^{i{2\pi\over N}}\phi_0$. 
Breaking this classical solution along the plane $x_0=0$ and rotating
the piece $x_0<0$ by $2\pi/N$ produces precisely the configuration
with the correct boundary conditions and the discontinuity structure.
The path integral in eq.(\ref{uexp}) is therefore dominated by this
classical configuration. Its action (up to corrections associated with
the boundary effects of $S$) is $\alpha S$ where $\alpha$ is the classical
wall tension of the domain wall which separates two adjacent $Z_N$ vacua. 
Thus we find that the expectation value of the U-loop is related to
the domain wall tension of the $Z_N$ domain wall by
\begin{equation}
<U(S)>=\exp\{-\alpha S\}
\end{equation}

\subsection{U-loop in the unbroken phase}

Now consider the unbroken phase. Again the U-loop average has the form of the overlap
of two states which factorizes as in eq.(\ref{fact}). Now however 
all observables noninvariant under $Z_N$  vanish in the vacuum. The action of the 
symmetry generator does not affect the state $|0>$. The state $|S>$ is therefore 
locally exactly the same as the state $|0>$ except along the boundary of the area $S$.
Therefore the only regions of space which contribute to the overlap are those
which lay within
one correlation length from the boundary. Thus 
\begin{equation}
<U(S)>=\exp\{-\gamma P(S)\}
\end{equation}
where $P(S)$ is the perimeter of the boundary of $S$.
The absence of the area law is again easily verified by a perturbative calculation.
In the unbroken phase the fluctuations of the field $\phi$ as well as the current density
$j_0=i (\pi\phi-\pi^*\phi^*)$ are small.
To leading order in the coupling constant
\begin{equation}
<U(S)>=\exp\big\{-
{1\over 2}\int _{x,y\in S}d^2xd^2y<j_0(x)j_0(y)>\Big\}
\label{sunbr}
\end{equation}
The possible area law contribution in the exponent is
\begin{equation}
S\int d^2x<j_0(0)j_0(x)>=S\lim_{p\rightarrow 0}G(p)
\label{area}
\end{equation}
where $G(p)$ is the Fourier transform of the charge density correlation function. 
The correlator of the charge densities however vanishes at zero momentum. 
This is because in the leading perturbative order the symmetry of the theory is 
actually $U(1)$ and not just $Z_N$ as seen in eq.~\ref{uexp}. Since the vacuum state is invariant it follows that 
the total charge $Q=\int d^2x j_0(x)=j_0(p=0)$ on this state vanishes, and
so does any correlation function that involves zero momentum component
of 
the charge density. So the area contribution in eq.(\ref{area}) is zero.
Strictly speaking in the leading order in perturbation theory eq.(\ref{sunbr}) 
is not the complete result. The exact expression contains in the exponential
also higher point correlators of the current density. Again however the possible 
area law contribution contains correlators of the total charge $Q$ with powers of $j_0$
and therefore vanishes. 

\subsection{U-loop  at high temperature}

Let us see now how the argument changes at high temperature.
The important difference is that the vacuum is not a pure state but rather a statistical
ensemble. 
The average 
of the U-loop is therefore not given by a single matrix element but rather by
\begin{equation}
<U>=\sum_ie^{-E_iT}<i|U|i>
\end{equation}
Let us consider the theory in which the $Z_N$ symmetry is broken at
zero temperature.
For concreteness we will think about $Z_2$ symmetric theory, although
qualitatively the discussion does not change for any $N$. 
The two degenerate vacuum states are characterized
by the value of the condensate $<\phi>=\pm\mu$.

In order to understand the behaviour of the $U$-loop we have to figure
out what
types of states contribute to the thermal ensemble.
At zero temperature the only states that are of interst are those with
finite energy. There are two towers of such states $|n>_\mu$ and
$|n>_{-\mu}$
- constructed above
each one of the degenerate vacua. These two towers of states 
do not talk to each
other, not only because their overlap is zero, but also because they
can not be connected to each other by action of any local (or semilocal)
operator  $_\mu<n|O|n'>_{-\mu}=0$. An immediate corollary of this is
that a superposition of the type
$|\alpha,\beta>=\alpha|n_\mu>+\beta|n>_{-\mu}$ violates
clustering property of the correlators of local operators.
For this reason at zero temperature 
in a spontaneously broken theory we are never interested in states
which carry sharp quantum numbers of the broken symmetry.

At finite temperature however we are also asking after states with
finite energy density, and therefore
infinite energy. This part of the spectrum looks rather different
if the energy density involved is high enough. The 
two vacuum configurations of the
potential
in eq.(\ref{lagr}) are separated by a finite barrier. Let us call the
hight of this barrier $H$.
The states with energy density lower than $H$ still separate into two
towers. We will denote these states by $|l>$.
However higher energy density states, with $\epsilon>H$ have different
nature. Their wave function
is not localized in the field space to the vicinity of one of the
vacua, but rather is spread over distances larger than the distance
between the two vacuum states $2\mu$.
These states therefore naturally
carry sharp quantum numbers with respect to the broken $Z_2$ symmetry.
These states we will denote by $|h>$.
In fact one expects that the higher the energy density the more
these states look like the
multiparticle states of a symmetric phase. 
That is to say as long as $\epsilon>>H$  
it does not matter whether the potential has the double well structure
or a single vacuum. These highly excited states
should look like states with finite density of particles which carry
the $Z_2$ charge.

At low temperatures, when the entropy effects are not important the
contribution to the thermal ensemble comes only from the $|l>$ -
sector since the Bolzman factor for any of the $|h>$ states vanishes 
exponentially in
the infinite volume limit.
As we have argued earlier, the average of $U$ in each
one of 
these states
has an area law behaviour and so does the whole temperature average of $U$. 

When the 
temperature reaches $T_C$ the phase transition occurs.
The reason for the onset of the phase transition is 
that when the equilibrium energy density
reaches critical
threshold value, the $|h>$ sector states start contributing to the
thermal ensemble. The sudden change in the entropy due to these new 
channels drives the phase 
transition. Above the phase transition therefore there are two kinds
of states that 
contribute to thermal averages.
One can 
then write
\begin{equation}
<U>_{T>T_c}=\sum_ne^{-E^l_n/T}<n,l|U|n,l>+\sum_se^{-E^h_n/T}<n,h|U|n,h>
\label{arper}
\end{equation}
where $n$ stands for all other quantum numbers.
In fact once the entropy effects become important enough to
excite the $|h>$-sector, the contribution
of $|l>$ states to any physical observable becomes negligible. 
As discussed earlier
each state in the second term gives a perimeter contribution
$\exp\{\gamma_sP\}$ to the average of the $U$-loop, so one could
be tempted to conclude that the loop must have a perimeter law
just like in zero temperature vacuum of a symmetric phase. This
however is not necessarily the case.
The reason is that $<n,h|U|n,h>$ is not positive definite. In fact
the number of states in which it is positive is roughly equal to the
number of states on which it is negative.
It is therefore very likely that the leading perimeter behaviour will
cancel and the net result will again be an area law for $<U>$.

Indeed if the ensemble can be thought of as an ensemble of
$Z_2$ charged free particles, the area law for $U$-loop follows
immediately\footnote{This argument is borrowed from \cite{mike}.We
thank Mike Teper and Biaggio Lucini for discussions of this point.}.
The $U$-loop in such an ensemble is
\begin{equation}
<U(S)>=\sum_{x_i,n}{1\over n!}\mu^n(-1)^n
\end{equation}
where $\mu$ is the fugacity of a single particle, and the summation goes over
all coordinates $x_i$ of the particles 
inside the area $S$ and over all possible
numbers of particles $n$. Assuming that particles have a finite size
$\Delta$, so that there are $S/\Delta$ possibilities to place one
particle
inside the area $S$ in the dilute gas approximation the sum gives
\begin{equation}
<U(S)>=\exp\{-{\mu\over \Delta}S\}
\label{us}
\end{equation}
We stress that the thermal ensemble of particles 
is $Z_2$ invariant. The density matrix
for such an ensemble can be written in the particle basis as
\begin{equation}
\rho=\sum_{n(x)}{1\over n!}\mu^n|n><n|
\end{equation}
The operator of the $Z_2$ transformation acts on the $n$-particle
states as
\begin{equation}
U|n>=(-1)^n|n>
\end{equation}
and so
\begin{equation}
U\rho U^\dagger=\rho
\end{equation}

The explicit simple formula eq.(\ref{us}) is derived in the dilute gas
approximation. We expect that the physics will be similar as long as
the interaction between the particles is short range.
Whenever the interaction is long range 
the behaviour of $<U(S)>$ can be different. 
For instance one does not expect area law behaviour
if the particles are bound into pairs since in this case only $Z_2$
invariant
states contribute to the thermal ensemble.

Our conclusion is that at finite temperature the behaviour of the
$U$-loop
is not strongly related to 
the mode of the realization 
of the $Z_N$ symmetry.
It is rather more likely to have an area behaviour.

To reiterate, the physics involved is very simple. At zero temperature
when acting on a
state,
the $U$-loop performs the $Z_N$ transformation inside the loop. The
only degrees of freedom that are changed by this operation inside the
loop,
are the $Z_N$ - noninvariant fields.  If the vacuum
 wavefunction depends on the configuration of
the noninvariant degrees of freedom (the state in question 
is not
$Z_N$ invariant) the action of $U$-loop affects the state everywhere
inside the loop. The VEV of $U$-loop then falls off as an area.
If the vacuum is $Z_N$ invariant, the wavefunction does not depend on
the configuration of the noninvariant degrees of freedom. The action
of $U$-loop then perturbs the state only along the perimeter, hence
the perimeter law in the unbroken phase.

At finite temperature however the thermal ensemble may even in 
the symmetric phase
contain significant contributions from states with nonvanishing $Z_N$
charges.
The $U$-loop therefore perturbs the thermal ensemble very
significantly everywhere inside the area, and the 
natural outcome is an area law.

The argument is quite general and does not depend on the exact form of 
the $Z_N$ invariant potential and more generally on the field content
of 
the theory - we 
could have added any number of extra fields to the theory
eq.(\ref{lagr}) without changing the conclusions.
The same relation must exist between the mode
of realization of the magnetic $Z_N$ symmetry and the behaviour of 
the Wilson loop
in the pure Yang-Mills theory.
The direct analogs of the scalar field $\phi(x)$ in eq.(\ref{lagr}) and 
the U-loop of the scalar theory are correspondingly 
the vortex field $V(x)$, and the 
spatial Wilson loop $W(C)$.

As we have shown in the previous section, the magnetic $Z_N$ is 
restored at high
temperature. The Wilson loop is nevertheless likely to have an area law
as is indeed indicated by all existing lattice data. In this context 
we note that analytic strong coupling results also give area behaviour\cite{borgs}.

\subsection{Wilson loop in 3+1 dimensions}

The previous considerations generalize to the 3+1 dimensons.  
At zero temperature in the broken phase
when acting with the Wilson loop $W(C)$ on the vacuum one changes the state of those 
magnetic vortices which loop through $C$. The number of such vortices which
are present in a generic configuration in the broken phase 
is proportional to the minimal area 
subtending $C$. 
The number of the degrees of freedom that is changed by the action of $W$
is thus proportional to the area $S$. Each of these degrees of freedom 
contributes a factor smaller than unity to 
the overlap with the vacuum state and so the VEV of $W$ scales with the 
exponential of the area.
In the unbroken phase the vacuum does not contain vortices of
arbitrarily large size.
The size of the vortices present in the vacuum 
is cutoff 
by the relevant correlation length. 
This is the case if the gauge group is completely broken by the Higgs
mechanism.  
Therefore for contours $C$ of linear dimension
much larger than this length, the action of
$W(C)$ only disturbs degrees of freedom close to the contour $C$ itself and
the VEV must have the perimeter behaviour.

At high temperature even the symmetric thermal ensemble is 
populated by vortices. These vortices are not free since
apart from them the ensemble also contains
``free'' charges. However unless this background of charges induces long range
interactions between the vortices, the most probable result for the
Wilson loop is the area law. 
A more detailed knowledge of vortex dynamics is necessary to
draw a firm conclusion.

To close this section we note that the present considerations
do not apply to Abelian theories. The magnetic symmetry does exist in this case too, but
here it is the continuos $U(1)$ 
group and the spectrum is massless. In this case there is no reason to
expect the local factorization of the overlap and generically therefore the arguments 
of this section
do not hold. In particular in the presence of long range correlations 
it is perfectly possible that the Wilson loop has a perimeter 
law even though
the state is perturbed everywhere inside the area boundede by the 
loop\footnote{In 2+1 
dimensions it is actualy only
the noncompact Abelian theories that are excluded from the consideration. Compact 
theories are massive and therefore should behave in the same way as the nonabelian 
Yang-Mills.}.

\section{Discussion.} 

In this paper our aim was to point out two facts. First that the calculation of the VEV
of the 't Hooft loop \cite{KAKS} implies the restoration of the magnetic $Z_N$ symmetry above
the deconfinement transition. Second, that the mode of realization of
magnetic symmetry is closely related to the
behaviour of the
spatial Wilson loop.

At zero temperature this relation is very rigid: spontaneous breaking
of $Z_N$ implies the area law behaviour for $W$ while unbroken $Z_N$
leads to perimeter law behaviour. 
At high temperature however even though the $Z_N$ symmetry is
restored, the Wilson loop may have an area law. This is 
the consequence of the fact that even a $Z_N$ - invariant
the thermal ensemble can contain a significant contribution of $Z_N$
nonsinglet states.
The area law is particularly simply understood
if the thermal ensemble at high temperature is well appproximated by
an ensemble of weakly interacting magnetic vortices.
The vortex gas argument has been 
previously brought up in favour
of the area behaviour of the spatial Wilson loop in
~\cite{shortvortex}.
This behaviour is also confirmed by
several lattice gauge theory
calculations \cite{karsh}. 

An alternative possibility is that 
due to the as yet unknown vortex dynamics,
there is vortex -  antivortex binding.
To explore such a possibility it would be very interesting
to measure on the lattice the free energy of a magnetic vortex.
In ref.(\cite{thooftperiodic}) the behaviour of the free energy of magnetic and
electric fluxes has been discussed 
in the {\it{low}} temperature phase.
To be able to do it in the lattice framework one has to define the theory
in a finite volume. As discussed by 't Hooft this can be 
achieved by imposing on the potentials
periodic boundary conditions modulo a gauge
transformation.
As discussed in ref.~\cite{thooftperiodic} this admits the presence
of vortices in 2+1 and of the vortex lines
in 3+1 dimensions.  
't Hooft's discussion was based on a Euclidean rotation
identity for the twisted 4d path integrals valid for any temperature,
and a factorization property of magnetic and electric fluxes.
In the notation of ref.\cite{thooftperiodic}:
\begin{equation}
F(\vec e,\vec m)=F_e(\vec e)+F_m(\vec m)
\end{equation}
Its validity  at low T is very reasonable, but is inconsistent 
with the Euclidean rotation identity at high T.
Based on this 't Hooft could prove ($N\leq
3)$ 
that in the confining phase, where the free energy of an
electric flux
is linear with the length (with the string tension $\rho$), 
the free energy of 
magnetic flux vanishes exponentially in the infinite volume limit. 
For a magnetic flux in,
say the z-direction it is
$\exp{-\rho L_xL_y}$. Thus the free energy of a magnetic flux
is related to the behaviour of the Wilson loop. The free energy of an electric
flux in the z direction in the {\it {hot}} phase vanishes exponentially like
$\exp{-\alpha L_xL_y}$ where $\alpha$ is the surface tension found in ref.\cite{KAKS}.
So the next obvious question is how the magnetic flux free energy behaves
in the hot phase. 
In a vortex gas picture this free energy should vanish exponentially
in the infinite volume limit. Such a calculation in 2+1 dimensions 
has been performed and (modulo some uncertainty related to imperfect
measurement of global vorticity) results are consistent with this
expectation \cite{mike}. It would also 
be instructive to see how in the hot phase
the additivity of electric and magnetic fluxes is broken.

We note that a recent lattice calculation \cite{rebbi} 
measures the monopole-antimonopole correlation. The results of 
\cite{rebbi} point to the screened behaviour of this correlation
function for {\it{all}} temperatures. So in the hot phase it behaves like  
its electric partner, the correlator of Polyakov loops\footnote{We 
note however that this simulation~\cite{rebbi} also points to the Coulomb
behaviour for the spatial 't Hooft loop in the hot phase, in
contradiction 
to analytic results~\cite{KAKS}\cite{pis} and early lattice 
results~\cite{kajantie}.  We feel that here more work
should be done to clarify the situation.}. 
This via 't Hoofts argument, is consistent with the measured
area behaviour of Wilson loops~\cite{karsh} and would imply that 
the magnetic flux free energy would fall off with an area law for
{\it{all}} 
temperatures. 

It is interesting to note that the vortex gas picture is equally applicable
in high temperature confining and nonconfining gauge theory. In
particular
one can consider an $SU(N)$ gauge theory with sufficient number of
adjoint Higgs fields, so that the gauge theory at zero temperature is
broken completely. In this situation the magnetic $Z_N$ symmetry is
unbroken
in the vacuum and the Wilson loop has a perimeter law. 
Magnetic vortices are finite energy excitations with
the mass of order $M=M_v^2/g^2$, where $M_v$ is the vector boson mass.
When the system is heated one expects that the thermal ensemble will
contain a dilute gas of these vortices at any temperature. 
Therefore at any finite temperature the spatial string
tension should be nonzero, although at low temperatures it will be
exponentially suppressed if the theory is weakly coupled:
$\sigma\propto M_v^2\exp\{-M_v^2/g^2T\}$.

{\bf Acknowledgements}
The work of A.K. is supported by PPARC. The work of CPKA and AK is
supported by a joint CNRS-Royal Society
project. We are indebted to Frithjof Karsch for bringing ref.\cite{borgs} to our attention. We thank Ian Kogan, Biaggio Lucini and Mike Teper for 
very interesting and
useful dicsussions, which in particular helped to clarify a
fundamental flaw in the arguments in the first version of this paper.

\pagebreak

\end{document}